\documentclass[aip,rsi,reprint]{revtex4-1}
\usepackage{graphicx}
\usepackage{amsmath,amsmath}
\usepackage{dcolumn}
\usepackage{bm}

\begin{document}


\title[Long working distance objective lenses for single atom trapping and imaging]{Long working distance objective lenses for single atom trapping and imaging}

\author{J. D. Pritchard}  \email{jonathan.pritchard@strath.ac.uk}
\affiliation{Department of Physics, University of Wisconsin-Madison, 1150 University Avenue, Madison, Wisconsin 53706}
\affiliation{Department of Physics, University of Strathclyde, 107 Rottenrow East, Glasgow, G4 0NG, UK}%
\author{J. A. Isaacs}
\affiliation{Department of Physics, University of Wisconsin-Madison, 1150 University Avenue, Madison, Wisconsin 53706}
\author{M. Saffman}  
\affiliation{Department of Physics, University of Wisconsin-Madison, 1150 University Avenue, Madison, Wisconsin 53706}

\date{\today}

\begin{abstract}
We present a pair of optimized objective lenses with long working distances of 117~mm and 65~mm respectively that offer diffraction limited performance for both Cs and Rb wavelengths when imaging through standard vacuum windows. The designs utilise standard catalog lens elements to provide a simple and cost-effective solution. Objective 1 provides $\mathrm{NA}=0.175$ offering 3~$\mu$m resolution whilst objective 2 is optimized for high collection efficiency with $\mathrm{NA}=0.29$ and 1.8~$\mu$m resolution. This flexible design can be further extended for use at shorter wavelengths by simply re-optimising the lens separations.
\end{abstract}


\maketitle

\section{Introduction}

High resolution imaging plays a crucial role in atomic physics experiments, from trapping and imaging single atoms \cite{schlosser01,sherson10,piotrowicz13,nogrette14} to creating complex arbitrary optical potentials for quantum gas studies \cite{ramanathan11,zimmermann11,gaunt12}. These experiments are performed under vacuum requiring imaging through a glass window  which introduces significant spherical aberrations and substantially reducing the resolution of an uncompensated optical system. One approach is to use lenses mounted inside the vacuum system such that collimated light passes through the viewport \cite{sortais07,tey08} however adjustment of the optical system is not possible once under vacuum and can additionally limit the lowest achievable pressure. 

\begin{figure}[t]
\includegraphics{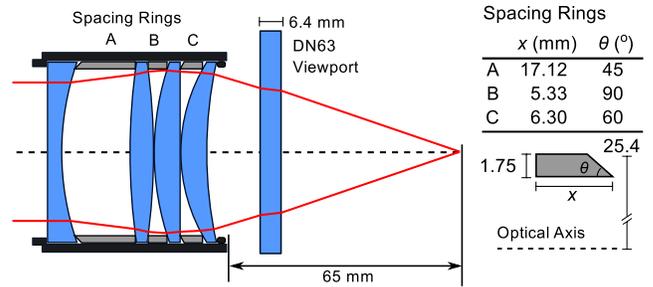}
\caption{(Color online). ATOM objective, optimized for high NA fluorescence detection for single atom readout at 852~nm.\label{fig1}}
\end{figure}

An alternative is to design a compound objective lens with the vacuum window included as the final element as first demonstrated by W. Alt \cite{alt02} which requires four singlet lenses to provide diffraction limited performance. A number of similar designs have since been published, either requiring one or more custom optics to be fabricated \cite{bucker09,zimmermann11} or using catalog optics but a short focal length ($\lesssim35$~mm) optimal for use with a small glass vacuum cell \cite{bennie13}. This precludes use with stainless steel vacuum systems that typically have much larger working distances.

In this paper we present a pair of optimized objective lenses with long working distances compatible for use with a standard stainless steel chamber. These are designed to be diffraction limited when used with standard conflat viewports and are comprized of standard catalog optics available with a range of anti-reflection coatings to provide a simple and cost-effective solution. Whilst designed for use at 802 and 852~nm for trapping and imaging neutral Cs atoms, the lenses remain diffraction limited at 780~nm for use with Rb without modification. For other elements, the lenses can be adapted to shorter or longer wavelengths by simply re-optimising the relative spacings.
\begin{figure*}[ht!]
\includegraphics{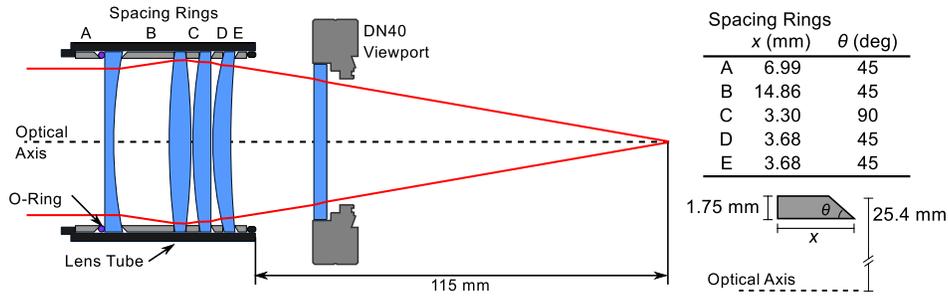}
\caption{(Color online). TRAP objective, optimized for diffraction limited performance at 802~nm.\label{fig2}}
\end{figure*}

\vspace{-0.1cm}

\section{Design}

The objectives presented in this paper are designed to provide infinity-corrected and diffraction limited performance whilst maximising the available numerical aperture given the minimum working distances imposed by our vacuum chamber geometry. For simplicity all elements are constrained to be 2~inch diameter BK7 singlets. The objectives are iteratively designed using ray tracing software (Zemax) to numerically optimise the lens spacing and curvatures to minimise the RMS wavefront error. One by one each element is then replaced by the closest matching catalog singlet and the remaining elements curvature and separation re-optimized until convergence is obtained. Interestingly, the objective designs converge to the same lens shapes as used originally by W. Alt \cite{alt02} suggesting this provides the optimal aberration compensation when the final element is a flat glass window. The objectives are then assembled in a standard 2 inch long lens tube (Thorlabs SM2L02) using spacing rings machined from delrin to avoid damaging the lenses.

The first objective (`ATOM') is designed to provide high efficiency fluorescence collection for imaging a single Cs atom at 852~nm through a standard CF4.5" fused silica viewport (thickness 6.4~mm) with a minimal working distance of 65~mm, giving a maximum achievable numerical aperture 0.3. Figure~\ref{fig1} shows the final lens configuration along with spacing ring parameters, with the complete prescription given in Table~\ref{ATOM}. This lens has an effective focal length of 67.4~mm and $\mathrm{NA}=0.291$ offering a collection efficiency of 2.1~\% and a diffraction limited modulation transfer function (MTF) on axis with an Airy radius of $r_\mathrm{A}=1.8~\mu$m. 

\begin{table}[t]
\begin{ruledtabular}
\begin{tabular}{c c c c}
Surface & Curvature (mm) & Thickness (mm) & Material \\
\hline
1 & $\infty$ & 4 & BK7\\
2 & 77.19 & 19.235 & Air \\
3 & 179.14 & 6.62 & BK7\\
4 & -179.14 & 0.2 & Air \\
5 & 77.26 & 7.29 & BK7\\
6 & $\infty$ & 0.2 & Air\\
7 & 47.87 & 7.29 & BK7\\
8 & 119.32 & 14.66 & Air\\
9 & $\infty$ & 6.4 & Silica\\
10 & $\infty$ & 49.2 & Vacuum
\end{tabular}
\end{ruledtabular}
\caption{ATOM Objective prescription - lenses from left to right are Thorlabs LC1611, LB1607, LA1417 and LE1418. The effective focal length at 852~nm is 67.4 mm.\label{ATOM}}
\end{table}

\begin{table}[t]
\begin{ruledtabular}
\begin{tabular}{c c c c}
Surface & Curvature (mm) & Thickness (mm) & Material \\
\hline
1 & $\infty$ & 2.5 & BK7\\
2 & 129.2 & 15.6 & Air\\
3 & 205.0 & 6.2 & BK7\\
4 & -205.0	 & 0.5 & Air\\
5 & 154.5 & 5.1 & BK7\\
6 & $\infty$ & 0.5 & Air\\
7 & 100.1 & 5.1&BK7\\
8 & 279.1 & 20.8 & Air\\
9 & $\infty$ & 3.3 & Silica\\
10 & $\infty$ & 98.0 & Vacuum	
\end{tabular}
\end{ruledtabular}
\label{TRAP}
\caption{TRAP Objective prescription - lenses from left to right are Newport KPC067, Thorlabs LB1199, LA1256 and LE1985. The effective focal length at 802~nm is 120 mm.}
\end{table}

The second objective (`TRAP') is designed for use with light at 802~nm to create an optical trapping potential. Here the lens is used in conjuction with a CF2.75" fused silica viewport (thickness 3.3~mm) with a working distance $\ge110$~mm and numerical aperture limited by the internal diameter of the viewport to $\mathrm{NA}=0.17$. The optimized configuration is shown in Fig.~\ref{fig2}, with complete prescription in Table~\ref{TRAP}. This design has an effective focal length of 119.5~mm and $\mathrm{NA}=0.172$, resulting in a an Airy radius of $2.85~\mu$m. 

\begin{figure}[b]
\includegraphics{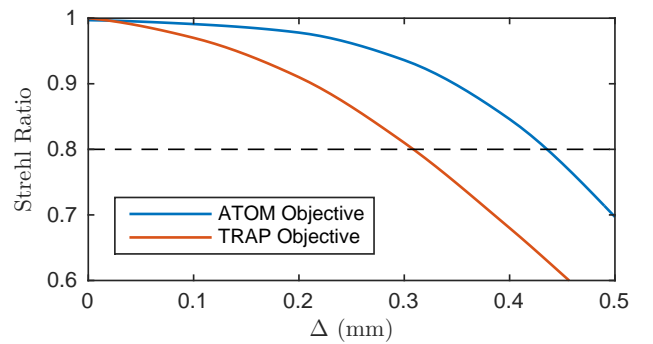}
\caption{(Color online). Calculated Strehl ratio $S$ for objectives. Diffraction limit defined as $S\ge0.8$.}
\label{fig3}
\end{figure}

In addition to considering the on-axis performance of the objectives, it is also possible to extract the diffraction-limited field of view (FOV). This is defined as the distance from the axis in the image space ($\Delta$) for which the Strehl ratio $S\ge0.8$ \cite{gross07}, and is plotted in Fig.~\ref{fig3} giving FOV diameters of $0.87$ and $0.62$~mm respectively.

Since the objectives are comprized of a single glass type they suffer from chromatic aberration. However, for small departures from the design wavelength no degradation in performance is observed, only a slight change in the effective focal length. For example, the ATOM objective at 780~nm has an effective focal length of 67.2~mm whilst at 852~nm the TRAP objective changes to 119.75~mm with a slight increase in $\mathrm{NA}$ to 0.175. Larger wavelength changes require re-optimisation of the lens separation, and this has been successfully applied to the TRAP objective to obtain diffraction-limited performance at 410~nm for imaging Ho atoms. 

\newpage

\section{Performance}

\begin{figure}[b]
\includegraphics{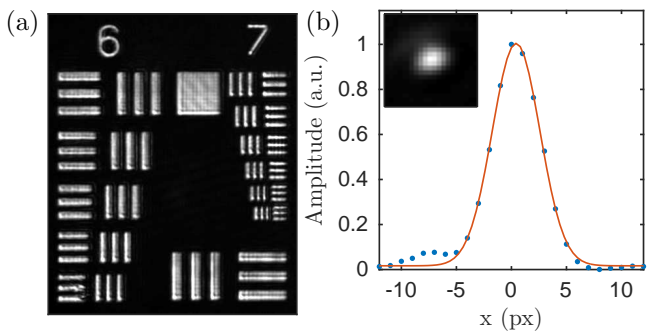}
\caption{(Color online). Calibration of imaging magnification and resolution for the ATOM objective using (a) 1951 USAF test card and (b) a 5~$\mu$m pinhole. Camera pixels are $8\times8~\mu$m.}
\label{fig4}
\end{figure}

To characterise the performance of the objective lenses, they are tested \textit{ex vacuo} using catalog AR coated fused-silica viewports. A relay telescope comprised of the objective lens in conjunction with an $f=1~$m singlet is used to image a USAF 1951 resolution target onto a CCD camera (Andor Luca R) using light at 852~nm. From these images the magnification and resolution are extracted by fitting the observed pattern to a square wave convolved with a gaussian point-spread function (PSF) characterised by standard deviation $\sigma$. Figure~\ref{fig4}(a) shows the test card image for the ATOM objective giving magnification $M=-14.9\pm0.1$ and $\sigma=0.8\pm0.1~\mu$m respectively, in excellent agreement with simulation ($M=-15$, $\sigma=0.62~\mu$m). For the TRAP objective $M=-8.4$ and $\sigma=1.8\pm0.2~\mu$m ($M=-8.36$, $\sigma=1.1~\mu$m). This large discrepancy in the PSF arises due to the imperfections in the optical quality of the viewport, as the full clear aperture of the window is used with substantial reduction in the optical flatness around the edges of the glass to metal seal. Improved results are possible either by post-selecting viewports or using home-made viewports from optical flats \cite{weatherill09,gupta12}. This does not affect the ATOM objective as this uses a larger viewport. Additionally, the ATOM objective is used to image a 5~$\mu$m pinhole which can again be treated as a top-hat profile convolved with the point spread function yielding $\sigma=0.75\pm0.05~\mu$m as shown in \ref{fig4}(b). 

\begin{figure}[t]
\includegraphics{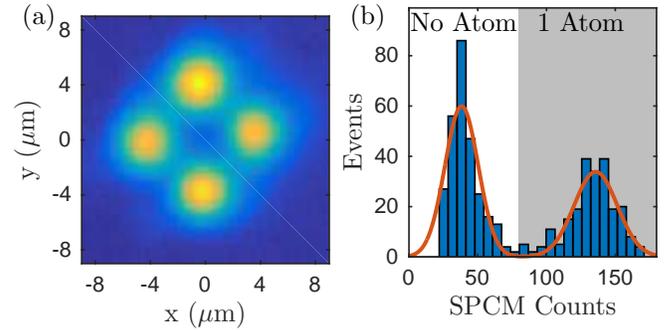}
\caption{(Color online). (a) Single-atom trap configuration using TRAP objective to create a single site GBA trap with four Gaussian spots with 3~$\mu$m waist separated by 6~$\mu$m. (b) Single atom fluorescence readout using the ATOM objective to demonstrate atom number discrimination, solid line shows fit to binomial distribution.}
\label{fig5}
\end{figure}

The objectives have subsequently been used for their designed application. Figure~\ref{fig5}(a) shows an image of the 802~nm trap configuration created using the TRAP objective to obtain with diffraction limited waists of 3~$\mu$m and 6~$\mu$m separation created using the Gaussian beam array (GBA) technique \cite{piotrowicz13}. This is achieved by combining a pair of frequency-shifted beams with orthogonal polarisations on a polarising beam splitter, then rotating the polarisation by 45 degrees and using a calcite to displace the beams to create four spots of equal intensity where each focus is a different frequency and polarisation from its neighbour to avoid interference effects at the centre of the beams. This creates a blue-detuned trapping potential for neutral Cs atoms which are localized within the saddle point at the centre of trap, with sufficiently small volume to enable single atom loading \cite{schlosser02} and trapping of both ground and Rydberg states \cite{zhang11}, ideal for application in neutral atom quantum computing \cite{saffman10}. For a total power of 350~mW, this gives a trap depth of $U_0=750~\mu$K and radial (axial) trap frequencies of $\omega_{r,z}/2\pi=25,1.5$~kHz.

Figure~\ref{fig5}(b) shows single atom readout measurements performed using the ATOM objective to provide high collection efficiency (2.2\%). The atom is trapped in a $2.4~\mu$m red-detuned dipole trap at 935~nm focused through the objective lens using a convergent input beam to compensate the 205~$\mu$m chromatic shift and overlap the 935~nm waist with the 852~nm focal plane. The atom is loaded from a standard magneto-optical trap using light assisted collisions to ensure either 0 or 1 atom in the trap \cite{schlosser02}. Readout is performed using light at 852~nm detuned by $\Delta/\Gamma=-4.4$ from the $6S_{1/2}~F=4\rightarrow6P_{3/2}~F'=5$ cooling transition with a total intensity of 10~mW/cm$^2$, corresponding to a scattering rate of 750~kHz. To overcome the differential AC Starkshift in the trap, the cooling light is chopped out of phase with the trap light at 1~MHz with a 50~\% duty cycle. The collected fluorescence is then collimated by the objective and re-imaged into a 1550~nm single mode fiber connected to a single photon counting module (Perkin Elmer SPCM-ARQH-13) with quantum efficiency of 45~\% at 852~nm. Transmission through the optical system (including transmission losses from dichroic optics and fiber coupling) is calibrated at 30~\%, giving an expected single atom signal of $\sim1$~counts/ms in excellent agreement with the observed histogram data which shows clear discrimination between 0 and 1 atom in the trap with an average signal of 100 counts above background and demonstrating single atom resolution with the ATOM objective.

\section{Conclusion}
We have presented a pair of objective lenses offering diffraction-limited performance and long-working distance for use with standard optical viewports suitable for single atom trapping and readout. The objectives are highly versatile, using standard catalog optics to provide a simple and economical approach to obtaining high resolution objective lenses. The design can be adapted to work over a range of wavelengths and can be used at multiple wavelengths simultaneously if the convergence of the secondary wavelength is adjusted to compensate the chromatic shifts as demonstrated above. Zemax lens files for the two objectives are included as supplemental material\cite{supp}.

\begin{acknowledgments}
This work was supported by funding from the NSF award PHY-1212448 and the University of Wisconsin Graduate School.
\end{acknowledgments}

\end{document}